\documentclass[lettersize,journal]{IEEEtran}
\usepackage{amsmath,amsfonts}
\usepackage{algorithmic}
\usepackage{algorithm}
\usepackage{array}
\usepackage[caption=false,font=normalsize,labelfont=sf,textfont=sf]{subfig}
\usepackage{textcomp}
\usepackage{stfloats}
\usepackage{url}
\usepackage{verbatim}
\usepackage{graphicx}
\usepackage{cite}
\usepackage{xcolor}
\usepackage{soul}

\begin{document}

\title{CEF: Connecting Elaborate Federal QKD Networks}
% Connecting Elaborated Federal QKD Networks
% CEF-QCI: Coordinating the Execution of Federated Quantum Communication Infrastructures
% noi oferim o arhitectura de orchestrare a conectivitatii de retele elaborate QKD

\author{Alin-Bogdan Popa and Pantelimon George Popescu ~\IEEEmembership{}
        % <-this % stops a space
\thanks{Corresponding author: P.G.Popescu}
\thanks{A.B.Popa and P.G.Popescu are with National University of Science and Technology POLITEHNICA Bucharest (emails: \underline{alin\_bogdan.popa@upb.ro}, \underline{pgpopescu@upb.ro})}% <-this % stops a space
\thanks{Manuscript received XXXXX; revised YYYYY.}}

% The paper headers
\markboth{Journal of \LaTeX\ Class Files,~Vol.~14, No.~8, August~2021}%
{Shell \MakeLowercase{\textit{et al.}}: A Sample Article Using IEEEtran.cls for IEEE Journals}

\IEEEpubid{0000--0000/00\$00.00~\copyright~2021 IEEE}
% Remember, if you use this you must call \IEEEpubidadjcol in the second
% column for its text to clear the IEEEpubid mark.

\maketitle

\begin{abstract}
As QKD infrastructure becomes increasingly complex while being developed by different actors (typically national governments), interconnecting them into a federated network of very elaborate sub-networks that maintain a high degree of autonomy will pose unique challenges. We identify several such challenges and propose a 4-step orchestration framework to address them based on centralized research, target network planning, optimal QKD design, and protocol enforcement.
\end{abstract}


\section{Introduction}
\IEEEPARstart{W}{ithin} the developments in quantum technologies, quantum key distribution (QKD) holds a significant place. For one, it is a low-hanging fruit, since simple QKD protocols such as BB84 are widely understood and only require a single photon source, a single photon detector, and several inexpensive pieces of optical equipment which are common in optics laboratories. Secondly, with the advent of quantum computers (where, as of this writing, the record stands at more than 1000 qubits), it is only a matter of time until Shor's algorithm is feasible for encryption schemes widely used today such as RSA 2048. With estimates placing the "quantum singularity" event as early as (possibly) in 2027, severe danger looms with the "Harvest Now, Decrypt Later" (HNDL) strategy, especially considering governmental or military data which, in many cases, is expected to remain classified for more than 30 years. As QKD provides unconditionally-secure keys (that is, security which does not rely on assumptions of limited computational power of a potential adversary) guaranteed by the laws of physics, it is a clear avenue for secure communication.

It is no wonder, then, that huge national and international efforts are being done towards understanding QKD, assessing its impact, deploying large-scale QKD networks, and integrating it into the existing infrastructure for critical or highly-sensitive communication. Multiple countries have recently published QKD as part of their national technological development strategy; on an international level, partnerships have been formed for cross-border terrestrial or satellite-based QKD deployments; extensive funding has been put forward \cite{mckinsey2024quantum}: as of April 2024 there were at least 96 startups in quantum communications gathering a cumulative investment of \$1.2 billion, and public funding in QKD has reached \texteuro180 million only in the EU.

The technology, however, is far from mature. There is a limited number of vendors which offer commercially available QKD devices (several notable vendors: IDQuantique, ThinkQuantum, LuxQuanta, Toshiba, QTI, QuintessenceLabs, Quantum CTek, KEEQuant, Quantum Industries); devices are often very sensitive to various environmental factors (temperature, movement and mechanical vibrations, optical fiber quality, power supply fluctuations, ambient light interference, and more) and even a minor variation results in service disruption; the key rates provided (especially in long-distance communications and at channel losses of more than 18 dB) are typically in the range of 1-4 kb/s, thus making them impractical for real-time unconditionally secure communication where the bit length of the key must match the bit length of the message to be transmitted; the QKD devices themselves are also quite expensive (typically around \$150,000) and the range of a single device is typically less than 165 km, thus making long-distance communications (which require multiple links) particularly difficult from a financial point of view; the infrastructure requirement also takes a toll: fiber optics links require dark fiber to be installed and maintained; terrestrial free-space links are generally only suitable on high-rise buildings and short ranges, and require high maintenance for optical alignment; satellite-based links can achieve longer range (i.e. intercontinental) key exchanges, but the initial cost to deploy a QKD-enabled satellite can exceed \$30 million.

Let us leave aside the QKD equipment and underlying infrastructure, and let us focus on the rest of the software stack, which is perhaps even more in its infancy. Although several standardization initiatives are under development (notably by ETSI and ITU-T), many essential components of a mature QKD internet are either now being defined or entirely unknown: high-level QKD network architectures; policies for security, access levels, and user clearance; protocols for horizontal interoperability between key management systems (KMSs); and many more.

\IEEEpubidadjcol

\begin{figure*}[t]
\centering
\includegraphics[width=\textwidth]{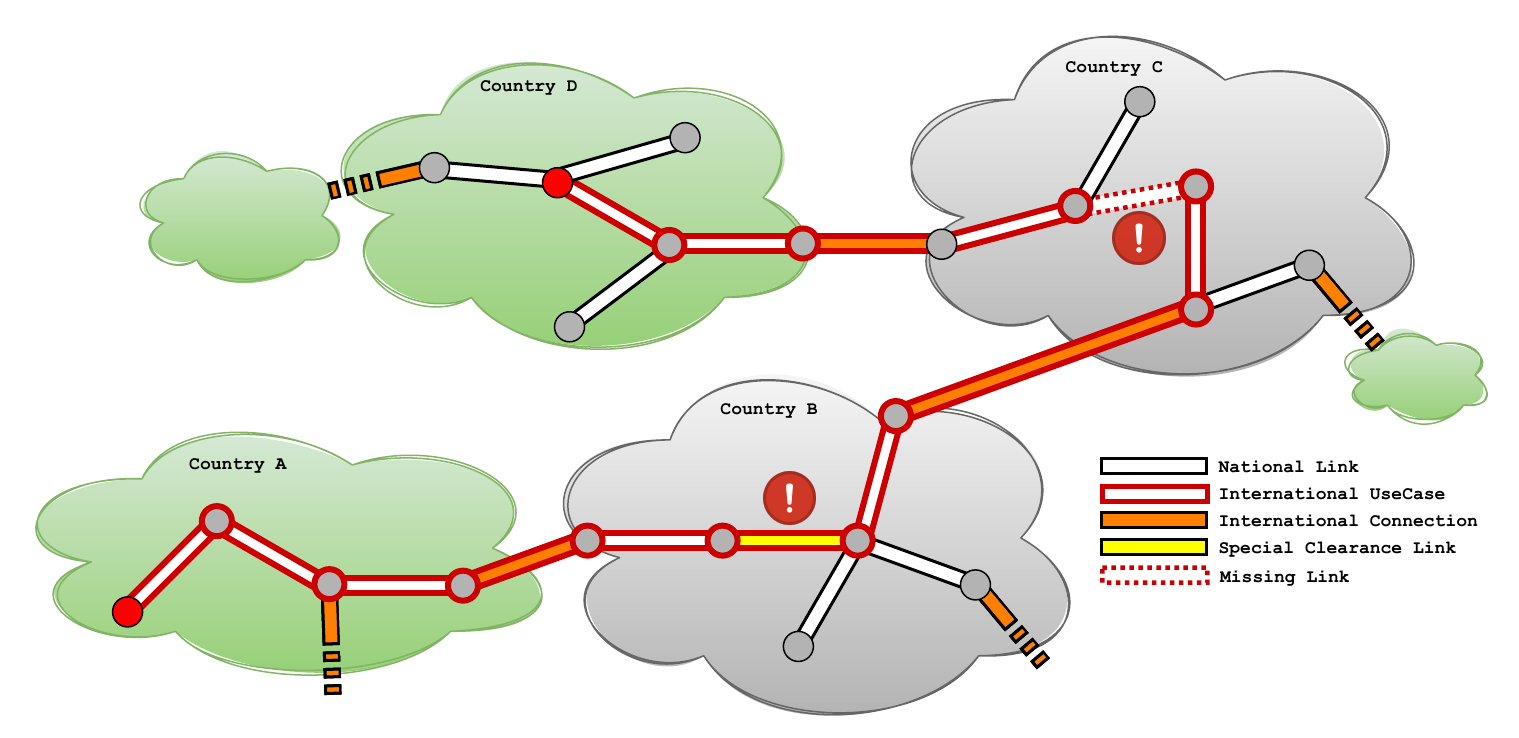}
\caption{Example of connected elaborate federal networks with connectivity issues that are due to the lack of centralized planning. With white are displayed national links within different countries, while international connections are orange. In this example, an international use-case is desired between the two nodes highlighted in red in countries A and D, with the link path that connects them highlighted in brown. While at first sight a path seems to exist (if networks B and C are considered black boxes), in reality two separate issues make this use-case impossible. First of all, the yellow link may connect two military sites, and requires special clearance to exchange keys with the rest of the network (which countries A and D do not have). Secondly, while country C does have international connections to both countries B and D, these two connection points are not actually connected with each other. The use-case between A and D is, unfortunately, not possible.}
\label{fig:stopcef_problems}
\end{figure*}

In the European Union, incredible progress has been done towards training experts, thoroughly testing the technology, as well as advancing the research in the problems defined above, through the European Quantum Communication Infrastructure (EuroQCI) initiative \cite{euroqci}. EuroQCI was started in 2019 and subsequently joined by all 27 EU Member States, with the purpose of building an EU-level QKD network in order to safeguard sensitive data and critical infrastructure (governmental institutions, data centers, energy grids, hospitals, and more). The first phase of EuroQCI was funded by the European Commission through the Digital Europe programme with an EU budget of \texteuro90 million (co-financed at least 50\% with the national governments) to support each Member State to build a national quantum communication infrastructure (NatQCI), and is expected to be finalized by the end of 2025. The second phase of EuroQCI will put forward another \texteuro90 million from EU funds and will involve interconnecting the NatQCIs; the funding call, which will be part of the Connecting Europe Facility (CEF) programme managed by the European Health and Digital Executive Agency (HaDEA), is currently expected to be announced in September 2024.

Nevertheless, deploying an international, federated network is not an easy task \cite{quantuminternet}. In this manuscript, we give a brief overview of the potential issues that may arise when the deployment follows a naive approach, and we provide an architecture and guidelines that aim to resolve these issues. For the sake of generality, we assume the goal is deploying a federated international network which consists of many national networks under the control of their respective country which are to be interconnected; however, the issues and solutions can apply to any set of QKD networks managed by different entities which are to be interconnected, even if the networks and entities are within the same country and governed by the same legislation. Throughout this paper, we will call the resulting network a Federated Quantum Communication Infrastructure (FedQCI).

\section{Arguments against the Naive Approach}

We define as "naive" any approach to deployment with the key characteristic that it lacks central management and orchestration of the deployment. The countries that are part of the deployment are the sole responsibles for integrating with each other into a larger QKD network; as such, the protocols for cross-border communication between any two countries are at the discretion of the respective countries. We can also assume each national network has its specific use-cases, key rate needs, and deployment infrastructure (including, but not limited to, vendor, brand, and supported protocols of the QKD devices used), which do not necessarily match between countries.

\begin{figure*}[t]
\centering
\includegraphics[width=\textwidth]{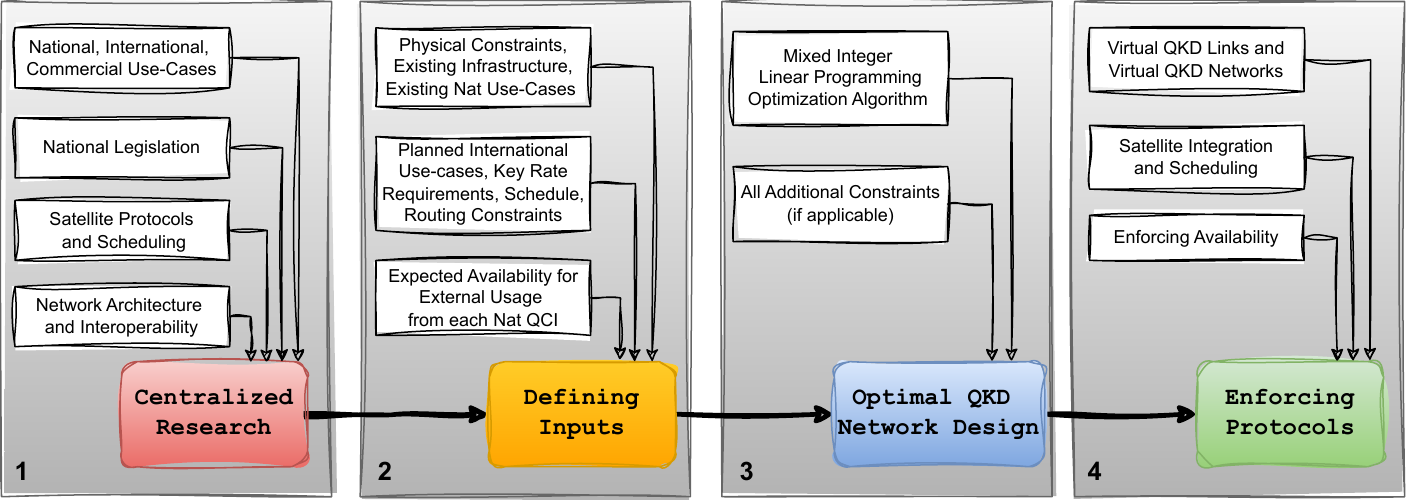}
\caption{Schematic of the proposed methodology, divided into 4 steps. Each step relies on the previous to have been completed.}
\label{fig:stopcef_proposal}
\end{figure*}

We start with the pressing \textbf{issue of ensuring minimal connectivity}, as displayed in figure \ref{fig:stopcef_problems}. It would be desirable for all partnering countries to have the guarantee that any-to-any connections are possible. However, without a centralized process for network planning, it is not inconceivable that the resulting network will not form a connected graph, essentially separating countries into independent clusters which cannot establish a secure communication session with other clusters. Special consideration must be paid to the countries which do not have multiple options for connecting to the rest of FedQCI. As a real-life example, Greece-Bulgaria-Romania-Hungary form a chain wherein each country relies on the next in order to achieve connectivity onwards to the rest of Europe, since it has no other neighboring countries that are also Member States. Additionally, countries that are islands or have significant island territories (Malta, Cyprus, Greece, Ireland) may need to rely strongly on satellite-based QKD to achieve connectivity, a requirement which is not as pressing for countries in Central and Western Europe. If the funding for a FedQCI is centrally managed, such differences in connectivity requirements must be taken into account for budget distribution.

A different issue is the \textbf{lack of standardization on security policies}. As national governments and policies change, so can the level of trust between neighboring intermediary countries which rely on one-another for long-distance QKD relaying. Security protocols have been proposed to mitigate this issue; for example, the centralized QKD KMS developed within OQTAVO-EuroQCI and published by Airbus Secure Communications \cite{airbuskms}, ensures no single malicious relay node can access the final key on a multi-hop link, by creating secure masks within each node that are processed in a central KMS held between the end-users (Alice and Bob). At minimum, a policy of this kind should be imposed on the international connections within a FedQCI.

Given the very limited key rates of commercially available QKD devices, it is probable that situations may arise where the \textbf{key requests between distant countries may clash with national QKD use-cases within intermediary countries}, due to missing clearance policies (one such example is also displayed in figure \ref{fig:stopcef_problems}). In recent research, we introduced the concept of Virtual QKD Networks (QKD VNets) \cite{futureqkd} as a solution to mitigate this issue by logically separating the key rates passing through a blackbox network into streams that can be treated as independent logical networks. Without consideration for this scenario, distant countries may often be deprived of access to secure communication with each other, and establishing access may only be possible through manual requests to all intermediary countries.

Considering the high sensitivity of QKD devices, relying on a single physical link to connect distant countries may not be ideal (consider, for example, a bottleneck link through which all secure keys are distributed between all western and eastern parts of Europe). Having \textbf{no consideration and preferential budget distribution for backup links} is, first of all, a recipe for frequent downtimes; and secondly, it is a low-hanging attack vector for malicious actors wishing to disrupt the international secure communication services.

Additionally, not considering all expected international use-cases from the get-go may lead to \textbf{sub-optimal use of the available budget}. By stating the requirements of the network in terms of key rate as a multi-commodity flow problem on a graph built following the available or feasible infrastructure and solving the problem with a Mixed Integer Linear Programming (MILP) solver, the cost optimality can be achieved while still ensuring all use-cases (or combination thereof) can be satisfied.

The \textbf{lack of protocols for horizontal KMS interoperability and software defined networks} is concerning. While some standardization initiatives are underway (notably: ETSI GS QKD series, standards 017, 018, 020, 021), they are not finalized and have not been publicly released. In the meantime, temporary protocols can be put in place to ensure interoperability during the testing phase, with clear policies on extending or upgrading the protocols when better standards get established. This is particularly important considering different NatQCIs are likely purchasing the QKD devices from different QKD vendors (which may have widely different supported features) or even develop their own in-house.

% it seems a strong emphasis within the upcoming CEF call for QKD is put on the development of optical ground stations (OGSs) for satellite-based communication through the upcoming Eagle-1 satellite. Eagle-1 is jointly developed by the European Space Agency (ESA) and the Luxembourg-based satellite company SES.

\begin{figure*}[t]
\centering
\includegraphics[width=\textwidth]{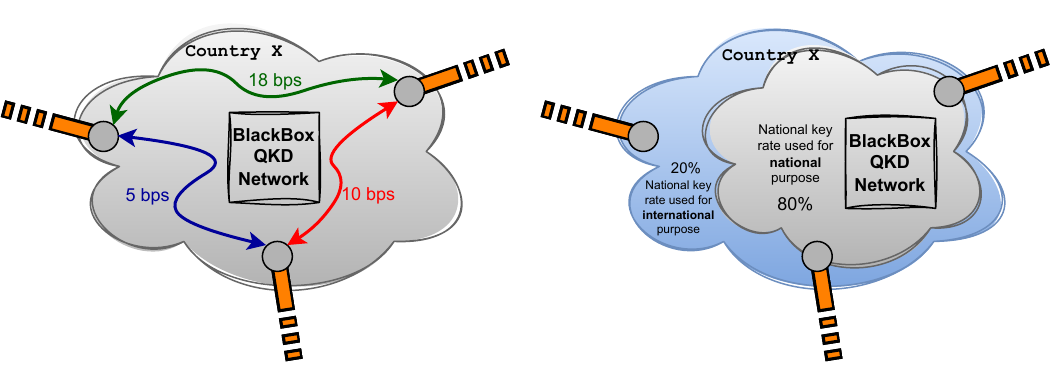}
\caption{On the left-hand side, Country X provides availability for external usage in the form of a guaranteed minimal key rate (in bits per second) between all pairs of external connections. On the right-hand side, Country X ensures 20\% of all its national QKD links are provided at the disposal of international use-cases.}
\label{fig:stopcef_availability}
\end{figure*}

Finally, the preference for terrestrial versus satellite-based connectivity needs to be discussed. At first glance, satellite links can enable QKD connectivity between very distant locations. However, \textbf{several issues with a satellite-first approach} can be noted: the deployment of an OGS is significantly more expensive than the most expensive terrestrial links (with typical prices starting at \texteuro1 million); if the satellites are not fully operational, countries investing in OGSs may not be able to actually use them, which may also lead to misalignment of the operational times of NatQCIs relying on satellite-based vs terrestrial-based links within the period where the international connections are to be maintained; the QKD specifications (for example, the expected key rate or frequency of periods of availability due to orbit scheduling) may be much lower for satellite links than for terrestrial fiber optics; more importantly there is a severe lack of protocols and policies for satellite-based QKD communication, which by nature is difficult to manage for a multitude of reasons: the satellite can only service one country at a time, the orbits have to be scheduled in advance, and the availability and expected key rate of the satellite depends on several factors including existing requests, current position, scheduled orbits, the weather, the time of day, and its commercial usage outside of the FedQCI.

\section{Architecture for a Successful FedQCI}

In the following we propose a 4-step methodology (see figure \ref{fig:stopcef_proposal}) which aims to mitigate most of the issues presented in the previous section.

The \textbf{first step} for a successful FedQCI is performing centrally managed research on key aspects involving the use-cases and software protocols involved within each country:

\begin{itemize}
    \item Research on actual national, international, and commercial use-cases, with the purpose of understanding how the beneficiaries of an international QKD deployment may desire to use the network. Can we narrow down whether the main users primarily consist of local authorities, national governing bodies, law enforcement agencies, European institutions, digital security agencies, military branches, hospitals, national emergency services, universities and research institutes, commercial computing clusters, banking services? Perhaps their needs are widely different. Is the number of nodes connected in the network more important, or is it the resulting key rate assuming the same network cost? Are there use-cases which have strong requirements for a specific number of nodes, a specific key rate, or a specific network layout? All these questions should be answered as part of this study.
    \item Research on the national legislations regarding information security and assess the feasibility of secure communication considering the definition and constraints for different levels of secrecy. Can country A actually use the QKD network to establish a secure communication session with country C, if the network passes through country B? Does it have specific requirements on the connection, such as a minimum level of security along every link or node? This study should not only answer these questions, but also get the national security bodies on the same page and expectations regarding QKD technologies.
    \item Research on satellite-based QKD protocols. How should a key request through the satellite look like? How should the satellite operator notify NatQCIs of scheduled downtimes or unavailability due to commercial usage? Can the satellite ensure a minimum key rate for every country that accesses it? How should the satellite availability be prioritized between a country with 1 OGS and a country with 5 OGSs, or between a small country which can be served on a single orbit and a large country which requires several orbits to fully service, or between countries with favorable versus unfavorable weather for satellite communications, or between an island country which must rely on the satellite and a landlocked country with plenty of terrestrial links at its disposal? How should the operators of an OGS schedule the satellite availability considering a multitude of requests may be in queue at the same time? If country A requests a key with country B via satellite, how is country B notified and what is the procedure for approval on behalf on country B? This research should work closely with the QKD networks desiring space connections and the operators of the satellite, and should be based on accurate simulations of satellite quantum communications.

    \begin{figure*}[t]
\centering
\includegraphics[width=\textwidth]{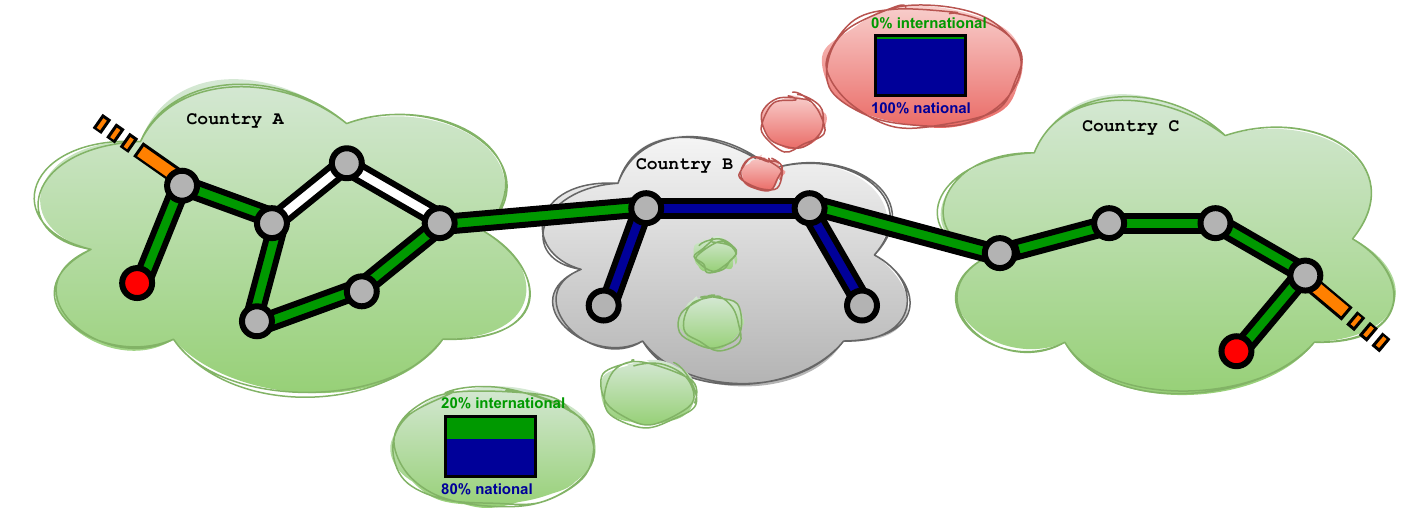}
\caption{An example of the effects of enforcing a minimum availability for external usage. Countries A and C desire to exchange QKD keys through country B for the use case delimited by the red nodes and highlighted in green. However, the link which passes through country B is used fully for a national use-case (highlighted in blue). Enforcing a minimum availability for international use-case (20\% in this example) would allow countries A and C to perform their use-case without a significant impact on B.}
\label{fig:stopcef_problemsolution}
\end{figure*}

    \item Research on network architecture and interoperability. Are the provisions of (future) standards such as ETSI GS QKD 020 enough? Depending on the expected use-cases, can we identify specific metadata that should become part of the inter-KMS communication? Should an operator in one country be able to remotely change the behaviour, the configuration, or the architecture of the QKD network in a different country, and if yes, when and how should this be possible? If a node, a link, a city, or perhaps an entire country gets compromised, how can keys stored in remote vaults or secrets based on compromised keys be made obsolete? How should the affected parties be notified? How should clearances be made for intermediary countries to allow distant countries to communicate securely? QKD VNets partly solve this last question, but the process of establishing VNets is still primarily manual unless proper policies are put into place and enforced. Linking international large-scale networks maintained by different actors comes with specific challenges, which should be addressed within this research. The resulting protocols and policies should be enforced onto the applicant projects.
    % \item Build the cost-optimal EuroQCI network design based on specific desired use-cases, connections, and key rates. Recent research shows how the constraints imposed on a network can be stated as linear equations and inequalities which can be solved for the optimal total network cost with a Mixed-Integer Linear Programming approach \hl{[cite morelinks care nu e gata..]}. This approach should be applied on the level of the entire EuroQCI and the budget should be distributed accordingly.
\end{itemize}

The \textbf{second step} is clearly defining the inputs to the design phase of the FedQCI:

\begin{itemize}
    \item \textbf{Physical constraints}: the list of partnering countries which are to be interconnected; their geographical position and their existing (or planned) infrastructure for fiber optics, QKD links, OGSs; a cost array with estimated costs for deploying a new QKD link or a new OGS for each location (the costs may not necessarily be the same across locations; for example, building an OGS from scratch in one city may be more expensive than upgrading an existing telescope to have QKD capabilities in a different city).
    \item \textbf{Planned international use-cases}: the list of end-to-end use-cases planned internationally (not necessarily between adjacent countries); their expected key rate requirements (if applicable), schedule (if not continuous), security requirements (e.g. one use-case may require all intermediary nodes to maintain a specific security level), and routing constraints (e.g. one use-case may require that keys do not pass through a specific country or region).
    \item \textbf{Availability for external usage}: the expected availability of each NatQCI for external use-cases (i.e. use-cases where the NatQCI in question is neither of the end users of said use-case). This availability may be defined in one of two ways (see figure \ref{fig:stopcef_availability}): as a point-to-point number of key bits per second between different cross-border nodes reserved for external use-cases, or as a percentage of all keys generated by the NatQCI. If the funding for the international deployment is provided by an independent international entity (such as the European Commission), perhaps a minimum availability from each NatQCI may be requested as part of the funding contract (perhaps proportional to the funds received and the existing national infrastructure from non-international funding). In the same way, we can define the availability for external usage for each cross-border link, as the key rate provided for use-cases which do not directly involve any of the two countries being part of the said cross-border link.
\end{itemize}

Given a clear definition of the inputs, the management entity is now ready for the \textbf{third step} which involves creating the international-level network design. The constraints imposed on a QKD network can be stated as linear equations and inequalities which can be solved optimally with a Mixed-Integer Linear Programming approach, hence, all use-cases, their schedules, the required and available key rates, and the availability of each country and cross-border terrestrial or satellite link can be taken into account to produce the theoretically-optimal network in terms of cost. If a different kind of optimality is desired (for example, maximizing the cross-country key rates for specific use cases, or maximizing the number of redundant links or routing options), then it can be applied within the same framework. If the international interconnection is centrally funded, this step also provides a clear budget distribution as per the specific requirements to the benefit of the international network.

The \textbf{fourth and final step} is enforcing the protocols needed for federated communication. At the minimum, each NatQCI should implement a VNet as described in \cite{futureqkd} such as to separate the national-level use-cases from the international key rates as per their specific agreed availability. If the FedQCI consists of countries that are to connect to the network via satellite, the satellite integration protocols produced during the research phase should be enforced at the level of the entire federated virtual network. It is also possible that adjacent countries (or chains of countries) may want to make use of the cross-border or satellite connections following a custom protocol; as such, all protocols should be enforced at the level of the federated virtual network only, which should not use up the entire available cross-country physical key rate (see figure \ref{fig:stopcef_problemsolution}). Practically, this can be implemented by setting a virtual sub-link within each cross-country connection, such that international-level connections can make use of the virtual FedQCI, leaving enough bandwidth for cross-country custom use-cases or research applications. 

\section{Conclusion}

Our suggestions can be summarized in the following: \textbf{we strongly recommend the international inter-connection of NatQCIs, and not only, to be centrally coordinated in terms of network design, communication protocols, research, and budget distribution.}

In this article we outline several issues identified with a naive decentralized approach to international QKD deployments, and we provide several suggestions which we hope will help resolve them. We expect any future large-scale international QKD deployment to face these challenges (including EuroQCI). On the plus side, we only need to solve these problems once. It is our firm belief that, with little centralized effort on research and management, the international QKD network can be successfully deployed and can have real-life use-cases outside of the research area, paving the way for integrating true secure communications on a global level.

\section*{Acknowledgement}

This work has been partially supported by RoNaQCI, part of EuroQCI, DIGITAL-2021-QCI-01-DEPLOY-NATIONAL, 101091562.

\section{References Section}
 
 % argument is your BibTeX string definitions and bibliography database(s)
%\bibliography{IEEEabrv,../bib/paper}
%

\section{Biography Section}
 
% \vspace{11pt}

\begin{IEEEbiography}[{\includegraphics[width=1in,height=1.25in,clip,keepaspectratio]{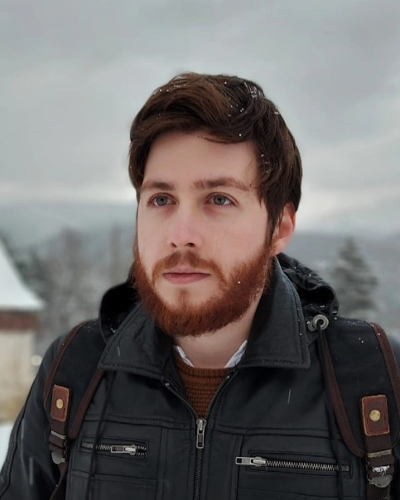}}]{Alin-Bogdan Popa} 
is a researcher at National University of Science and Technology POLITEHNICA Bucharest and a member of RoNaQCI (part of EuroQCI). He is CTO and partner at Parable, a venture studio and micro-to-small cap private equity fund with \$180M in assets-under-management. His main research interests are Quantum Communications, Decentralized Finance and Machine Learning. 
\end{IEEEbiography}

\begin{IEEEbiography}[{\includegraphics[width=1in,height=1.25in,clip,keepaspectratio]{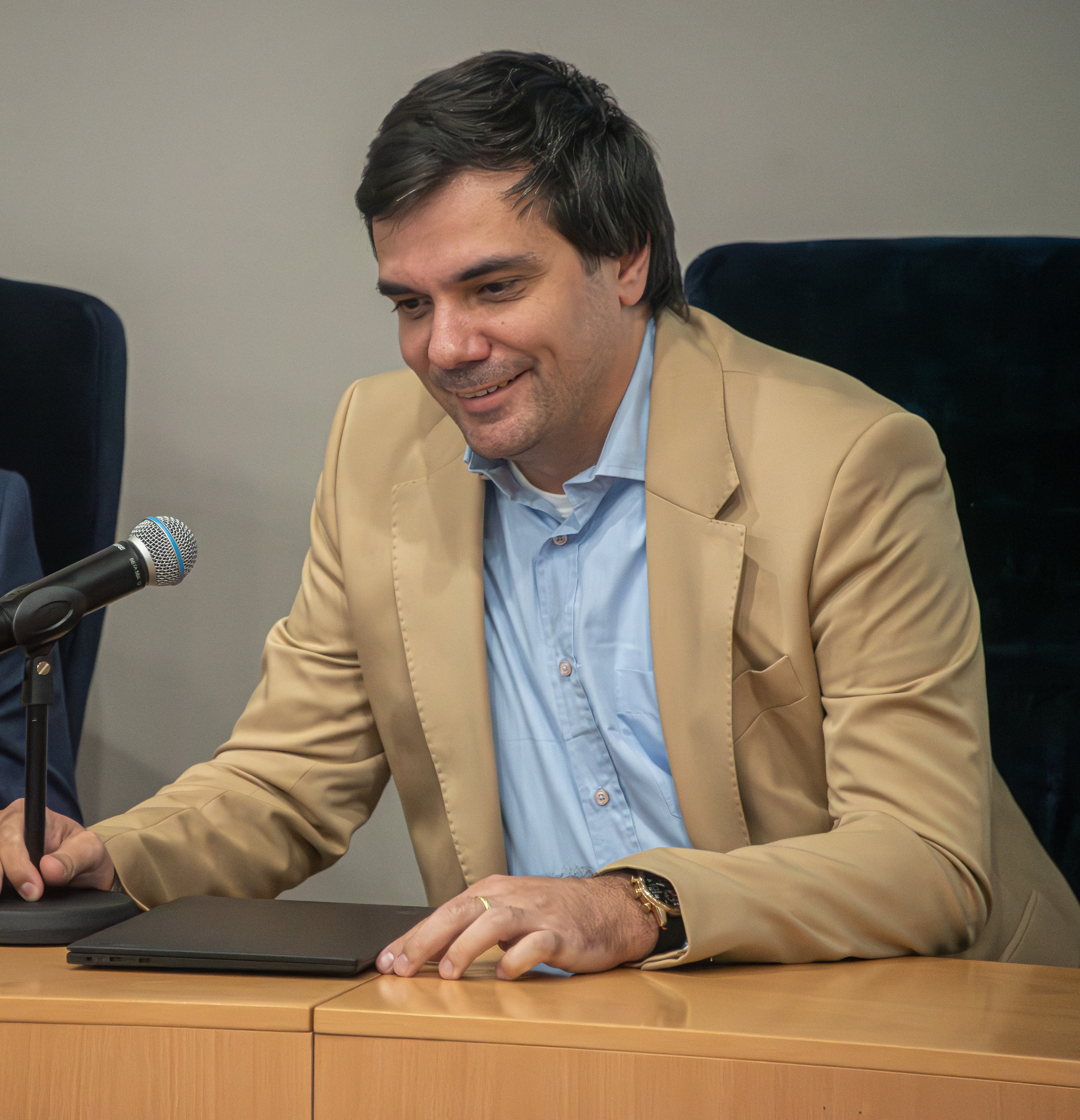}}]{Pantelimon George Popescu} is full Professor within the
Computer Science and Engineering Department and head of the Quantum Computing Laboratory at the National University of Science and Technology POLITEHNICA Bucharest. He is the Technical Coordinator of RoNaQCI (part of EuroQCI), member of the CERN QTI Advisory Board and member of the Open Quantum Institute Advisory Committee. His main fields of interest include Quantum Computing, Numerical Methods, Information Theory and Inequalities.
\end{IEEEbiography}

\vfill

\end{document}